\documentclass[twocolumn,showpacs]{revtex4}
\usepackage{psfig}
\begin{document}
\title{
{\boldmath$\omega$-$\phi$} mixing in chiral perturbation theory}


\author{Ayse Kucurkarslan$^{1}$, and Ulf-G. Mei{\ss}ner$^{2,3}$}

\affiliation{
$^1$ Canakkale 18 Mart University, 17020 Canakkale, Turkey\\
$^2$Helmholtz-Institut f\"ur Strahlen- und Kernphysik (Theorie), 
Universit\"at Bonn, Nu{\ss}allee 14-16, D-53115 Bonn, Germany \\
$^3$Institut f\"ur Kernphysik (Theorie), Forschungszentrum J\"ulich,
D-52425 J\"ulich, Germany.
}

\begin{abstract}
We investigate  $\omega$-$\phi$ meson mixing to leading order
in chiral perturbation theory utilizing the antisymmetric tensor
field formulation. We update the quark mass ratio $R$ from $\rho$-$\omega$
mixing, $R = 42 \pm 4$.
\vspace{-6.5cm}

\hfill {\tiny FZJ--IKP(TH)--2006--06, HISKP-TH-06/05}

\vspace{6.5cm}
\end{abstract}
\pacs{12.39.Fe; 12.40.Vv; 13.25.-k; }

\maketitle

\noindent {\bf 1.}
The mixing between the vector mesons  $\omega$ and $\phi$ plays an
important role in the understanding of OZI violation and SU(3) breaking
in QCD. In QCD, this mixing is entirely generated by the light quark 
mass differences. In addition, QED effects by photon exchange lead to
a further mixing contribution. In the framework of chiral perturbation
theory with explicit vector fields the issues related to $\rho$-$\omega$
mixing were worked out by Urech~\cite{Urech:1995ry}. In this short
note, we use the same framework to calculate  $\omega$-$\phi$ mixing
and also to discuss the momentum dependence of the mixing amplitude.
This might be of relevance for the theoretical studies trying to explain
the recently measured large $\phi$-meson photo-production cross section
of nuclei in the non-perturbative regime of QCD~\cite{Ishikawa:2005aw}.

\medskip

\noindent {\bf 2.}
We first give a brief reminder of $\omega$-$\phi$ mixing in the standard
quark model picture. Since there is some small OZI violation, 
one has an admixture of  light quarks in the $\phi$ meson wave function.
The $\phi$ and $\omega$ mesons are a mixture of the SU(3)
singlet $\omega_{0}$ and the octet $\omega_{8}$ states,
\begin{eqnarray}
\phi &=& \cos\theta_{V}\,\omega_{8}-\sin\theta_{V}\,\omega_{0} ~,\\
\omega &=&\sin\theta_{V}\,\omega_{8}+\cos\theta_{V}\,\omega_{0}~,
\end{eqnarray}
where
\begin{eqnarray}
\omega_{8}=\left(u\overline{u}+d\overline{d}-2s\overline{s}\right)/\sqrt{6}~,\\
\omega_{0}=\left(u\overline{u}+d\overline{d}+s\overline{s}\right)/\sqrt{3}~.
\end{eqnarray}
The $\phi$ and $\omega$ wave functions are then given by
\begin{eqnarray}
\phi &=& (u\overline{u}+d\overline{d})\left(\frac{1}{\sqrt{6}}\cos\theta_{V}-
\frac{1}{\sqrt{3}}\sin\theta_{V}\right)\nonumber\\
& &-s\overline{s}\left(\frac{2}{\sqrt{6}}\cos\theta_{V}+
\frac{1}{\sqrt{3}}\sin\theta_{V}\right)~, \\
\omega &=& (u\overline{u}+d\overline{d})\left(\frac{1}{\sqrt{6}}\sin\theta_{V}+
\frac{1}{\sqrt{3}}\cos\theta_{V}\right)\nonumber\\
&+& s\overline{s}\left(-\frac{2}{\sqrt{6}}\sin\theta_{V}+
\frac{1}{\sqrt{3}}\cos\theta_{V}\right)~.
\end{eqnarray}
The strict OZI limit  corresponds to the ideal mixing angle
\begin{equation}
\tan\theta_{V} =\frac{1}{\sqrt{2}} \to \theta_{V}=35.3^{\circ}~,
\end{equation}
and the ideal $\phi$ and $\omega$ states are
$\omega^{\rm ideal} = 2\omega_{0}/\sqrt{6} +\omega_{8}/\sqrt{3}\,$ and
$\phi^{\rm ideal}= \omega_{0}/\sqrt{3} - 2\omega_{8}/\sqrt{6}\,$.
It is further instructive to rewrite the $\phi$ and $\omega$ 
wave functions as follows
\begin{eqnarray}
\phi &=& \sin\varphi_{V}\left(\frac{1}{\sqrt{2}}(u\overline{u}
+d\overline{d})\right)-\cos\varphi_{V}(s\overline{s})\\
\omega &=& \sin\varphi_{V}(s\overline{s})+ \cos\varphi_{V}
\left(\frac{1}{\sqrt{2}}(u\overline{u}+d\overline{d})\right)~,
\end{eqnarray}
where $\varphi_{V}=\theta-\theta_{V}$. The physical mixing angle
$\theta$ can e.g. be determined from the masses of the mesons in the
vector meson nonet differs from the ideal mixing angle. Using
the quadratic Gell-Mann-Okubo mass formula, the physical mixing
angle of the vector mesons can be obtained as $\theta=39^{\circ}$, 
close to  the ideal one: $\varphi_{V}\simeq39^{\circ}-35.3^{\circ}=3.7^{\circ}$. 
Note further that the wave function of $\phi$ meson can also be written
as
\begin{equation}\label{eq:eps}
\phi(1020)=s\overline{s}
+\varepsilon_{\phi\omega}(u\overline{u}+d\overline{d})/\sqrt{2}
\end{equation}
where $\varepsilon_{\phi\omega}$ is the mixing parameter,
$\left|\varepsilon_{\phi\omega}\right|\ll1$, which describes the
$\phi\omega$ mixing amplitude. 
In what follows, we will determine the mixing amplitude and angle 
from vector meson decays.

\medskip

\noindent {\bf 3.}
For our analysis, we use the vector meson chiral effective Lagrangian
presented in~\cite{Ecker:1988te} and extended in~\cite{Urech:1995ry}. 
To construct the pertinent Lagrangian, 
we introduce the antisymmetric tensor field $V_{\mu\nu}$ to parameterize the 
octet of the spin-1 vector mesons
\begin{eqnarray}
V_{\mu\nu}=
     \left(%
\begin{array}{ccc}
  \frac{\rho^{0}}{\sqrt{2}}+\frac{\omega}{\sqrt{2}} & \rho^{+} & K^{*-} \\
  \rho^{-} & -\frac{\rho^{0}}{\sqrt{2}}+\frac{\omega}{\sqrt{2}} &  K^{*-0}\\
  K^{*-} & K^{*0} & \phi
\end{array}%
\right)_{\mu\nu}~.
\end{eqnarray}
To include the $\phi$ (the singlet field), we extend the 
$SU(3)$ representation to $U(3)$ and substitute
\begin{equation}
V_{\mu\nu}\rightarrow
V_{\mu\nu}+(\omega_{0})_{\mu\nu}\frac{I_{3}}{\sqrt{3}}~,
\end{equation}
where $\omega_{0}$ is the lightest singlet vector resonance and $I_3$
is the unit matrix in three dimensions. 
For the analysis of vector meson mixing, we need the
effective Lagrangian to leading order. The strong contribution 
to vector-meson mixing stems from the terms
quadratic in the vector meson fields, i.e. the 
kinetic part of the Lagrangian for the $\omega$ and $\phi$ mesons.
In the chiral limit, it takes the
form (from now on, we only display the terms needed for our discussion)
\begin{eqnarray} \label{e1}
{\cal
L}_{\rm kin}=-\frac{1}{2}\partial^{\lambda}\phi_{\lambda\mu}
\partial_{\nu}\phi^{\nu\mu}+\frac{1}{4}M^{2}_{\phi}
\phi_{\mu\nu}\phi^{\mu\nu}\nonumber \\
-\frac{1}{2}\partial^{\lambda}\omega_{\lambda\mu}
\partial_{\nu}\omega^{\nu\mu}+\frac{1}{4}M^{2}_{\omega}
\omega_{\mu\nu}\omega^{\mu\nu}\, .
\end{eqnarray}
The part of ${\cal L}_{\rm kin}$ that contains the $\phi-\omega$ mixing is
\begin{eqnarray} \label{e2}
{\cal
L}_{\phi\omega} = \sqrt{2}M_{\rho}(\widehat{m}-m_{s})\phi_{\mu\nu}
\omega^{\mu\nu}~,
\end{eqnarray}
where we have identified the octet mass in the chiral limit with the $\rho$ mass
as also done in~\cite{Urech:1995ry}. We further need the interaction 
Lagrangian,
\begin{equation}\label{Lint}
{\cal L}_{2}^{V}=
\frac{iG_{V}}{\sqrt{2}}\langle
V_{\mu\nu}u^{\mu} u^{\nu} \rangle +
\frac{F_{V}}{2\sqrt{2}}\langle V_{\mu\nu}
f_{+}^{\mu\nu}\rangle 
\end{equation}
where
\begin{eqnarray}
u_{\mu} &=&iu^{\dagger}D_{\mu}Uu^{\dagger}=u_{\mu}^{\dagger}~,\nonumber
\\
f_{+}^{\mu\nu}&=&uF^{\mu\nu}u^{\dagger}+u^{\dagger}F^{\mu\nu}u~,
\nonumber \\
F^{\mu\nu}&=&eQ(\partial^{\mu}A^{\nu}-\partial^{\nu}A^{\mu})~,
\end{eqnarray}
with $Q$ the quark charge matrix, $Q= {\rm diag}(2, -1, -1)/3$,
$U = u^2$ collects the Goldstone boson octet and $A_\mu$ is the
photon field.
The first term in Eq.~(\ref{Lint}) generates the vector meson couplings
to two pion which is needed for the calculation of the strong decay
channels.
The electromagnetic contribution to the vector meson mixing 
stems from the vector meson conversion
to the photon field and its reconversion into another neutral vector meson.
These terms are generated by the second term in Eq.~(\ref{Lint}).
Expanding this, the direct couplings of the  neutral vector meson fields
$(\rho^{0},\omega$, and $\phi)$ to the photon fields are given by
\begin{eqnarray} \label{e3}
{\cal L}_{\rho\gamma}=-\frac{e}{2}F_{V}\rho^{0}_{\mu\nu}
(\partial^{\mu}A^{\nu}-\partial^{\nu}A^{\mu}) ~, \nonumber\\
{\cal L}_{\omega\gamma}=-\frac{e}{6}F_{V}\omega_{\mu\nu}
(\partial^{\mu}A^{\nu}-\partial^{\nu}A^{\mu})~,
\nonumber\\
{\cal L}_{\phi\gamma}=\frac{e}{3\sqrt{2}}F_{V}\phi_{\mu\nu}
(\partial^{\mu}A^{\nu}-\partial^{\nu}A^{\mu})~.
\end{eqnarray}
The last two equations in Eqs.~(\ref{e3})
will lead to $\phi-\omega$ mixing through the
transition process $\phi-\gamma-\omega$. The corresponding
part of the Lagrangian takes the form
\begin{eqnarray} \label{e4}
{\cal L}_{\phi\omega}=
\frac{\sqrt{2}}{9}e^{2}F^{2}_{V}\phi_{\mu\nu}\omega^{\mu\nu}
\end{eqnarray}
Putting pieces together, the Lagrangian relevant for  $\phi-\omega$ mixing can be
written as (using Eqs.~(\ref{e2},\ref{e4}))
\begin{eqnarray}
{\cal
L}_{\phi\omega}=\Theta_{\phi\omega}\phi_{\mu\nu}\omega^{\mu\nu}
\end{eqnarray}
where the mixing angle $\Theta_{\phi\omega}$ has a strong and an
electromagnetic piece,
\begin{eqnarray}\label{Lphiom}
{\cal L }_2^{\phi\omega}=\left(\sqrt{2}M_{\rho}(\widehat{m}-m_{s})
+\frac{\sqrt{2}}{9}e^{2}F^{2}_{V}\right)\,\phi_{\mu\nu}\omega^{\mu\nu}~.
\end{eqnarray}
We remark that we can use here the average light quark mass 
$\widehat{m} = (m_u + m_d)/2$ since only its relative size compared
to the strange quark mass is of relevance.
Note that using  the lowest order expressions for the quark mass
expansion of the Goldstone boson masses,
we can rewrite the first term in Eq.~(\ref{Lphiom}) 
stemming from the quark mass differences 
entirely in terms of Goldstone boson masses. Further, we have employed
quark counting rules for the vector meson masses to leading order in the
quark masses. This leads to $M_V \simeq B_0/2$ (with $B_0 = |\langle 0 |
\bar qq|0\rangle|/F_\pi^2$) and lifts the apparent conflict with renormalization
group invariance of  Eq.~(\ref{Lphiom}) (for details, see~\cite{Urech:1995ry}).
%
Note further that the on-shell mixing amplitude $\Theta_{\phi\omega}$
is related to the parameter $\varepsilon_{\phi\omega}$ introduced in
Eq.~(\ref{eq:eps}) via
\begin{eqnarray}\label{defeps}
\varepsilon_{\phi\omega}
= \frac{\Theta_{\phi\omega}}{M^{2}_{\phi}-M^{2}_{\omega}}~.
\end{eqnarray}

\medskip

\noindent {\bf 4.}
To evaluate the mixing amplitude, we consider the Fourier transform of the
two-point function in the tensor field notation. It has the form
\begin{eqnarray} \label{amp1}
&i&  \!\!\!\!  \int d^{4}x \, e^{ikx}\langle
0|T\phi_{\mu\nu}(x)\omega_{\rho\sigma}(0)e^{i\int d^{4}y\{{\cal
L}_{2}^{V}\ + {\cal L}_{2}^{\phi\omega}\}} | 0 \rangle\nonumber \\
&=&\frac{2\sqrt{2}M_{\rho}(\widehat{m}-m_{s})}{M_{\phi}^{2}M_{\omega}^{2}}
\times\Biggl\{G_{\mu\nu\rho\sigma}\nonumber\\
&+&\biggl[
\frac{1}{M_{\phi}^{2}-k^{2}}+\frac{M_{\phi}^{2}}{(M_{\phi}^{2}
-k^{2})(M_{\omega}^{2}-k^{2})}\biggr]
P_{\mu\nu\rho\sigma}\Biggr\}\nonumber \\
&+&\frac{\sqrt{2}}{9}e^{2}F_{V}^{2}\frac{1}{(M_{\phi}^{2}
-k^{2})(M_{\omega}^{2}-k^{2})k^{2}}P_{\mu\nu\rho\sigma}~,
\end{eqnarray}
where
\begin{eqnarray*} 
G_{\mu\nu\rho\sigma} &=&  g_{\mu\rho}
g_{\nu\sigma}-g_{\mu\sigma}g_{\nu\rho}~,\\
P_{\mu\nu\rho\sigma} &=&g_{\mu\rho}k_{\nu}k_{\sigma}
-g_{\mu\sigma}k_{\nu}k_{\rho}-g_{\nu\rho}k_{\mu}k_{\sigma}+
g_{\nu\sigma}k_{\mu}k_{\rho}~.
\end{eqnarray*}
The on-shell amplitude is obtained in two steps. First, we must collect
some formula from  Ref.~\cite{Urech:1995ry} for $\rho$-$\omega$ mixing.
For that, consider  the decay width of the process $\omega\rightarrow\rho^{0}
\rightarrow\pi^{+}\pi^{-}$. It is expressed as
\begin{eqnarray}\label{ompipi}
\Gamma(\omega\rightarrow\pi^{+}\pi^{-})&=&\left|\frac{\Theta_{\rho\omega}}
{M^{2}_{\omega}-M^{2}_{\rho}-i(M_{\omega}\Gamma_{\omega}-
M_{\rho}\Gamma_{\rho})}\right|^{2}\nonumber\\ &\times&
\Gamma(\rho^{0}\rightarrow\pi^{+}\pi^{-})~,
\end{eqnarray}
and the decay width of $\rho^{0}\rightarrow\pi^{+}\pi^{-}$ has
been calculated using the antisymmetric tensor field Lagrangian in 
Ref.~\cite{Ecker:1988te},
\begin{eqnarray}
\Gamma(\rho^{0}\rightarrow\pi^{+}\pi^{-})
=\frac{1}{48\pi}\frac{G^{2}_{V}M^{3}_{\rho}}{F_{0}^{4}}
\left(1-\frac{4M^{2}_{\pi}}{M^{2}_{\rho}}\right)^{\frac{3}{2}}~,
\end{eqnarray}
with $F_0$ the pion decay constant in the chiral limit. In the numerical
analysis, we will identify this with the physical value of the pion 
decay constant, $F_0 = F_\pi = 92.4\,$MeV.
We further have~\cite{Urech:1995ry}
\begin{eqnarray}
\Theta_{\rho\omega}=2M_{\rho}(m_{u}-m_{d})+\frac{1}{3}e^{2}F^{2}_{V}~.
\end{eqnarray}
Of course, here we need to take care of the up and down quark 
mass difference since otherwise there would be no strong $\rho-\omega$ mixing.
Next we consider the decay width of the decay
$\phi\rightarrow\pi^{+}\pi^{-}$, we find
\begin{eqnarray}\label{phipipi}
\Gamma(\phi\rightarrow\pi^{+}\pi^{-}) &=& \left|\frac{\Theta_{\phi\omega}}
{M^{2}_{\phi}-M^{2}_{\omega}-i(M_{\phi}\Gamma_{\phi}-
M_{\omega}\Gamma_{\omega})}\right|^{2}\nonumber\\
&\times& \Gamma(\omega\rightarrow\pi^{+}\pi^{-})~,
\end{eqnarray}
where
\begin{eqnarray}
\Theta_{\phi\omega} =
\sqrt{2}M_{\rho}\left[(\widehat{m}-m_{s})+e^{2}\frac{F^{2}_{V}}{9M_{\rho}}\right]~.
\end{eqnarray}

\medskip

\noindent {\bf 5.}
We are now in the position to analyse the vector-meson mixings.
We use the following values $M_{\rho}=775.8\pm0.5\,$MeV,
$\Gamma_{\rho}=150.3\pm1.6\,$MeV, $M_{\omega}=782.59\pm0.11\,$MeV,
$\Gamma_{\omega}=8.49\pm0.08\,$MeV,
$M_{\phi}=1019.456\pm0.020\,$MeV, $\Gamma_{\phi}=4.26\pm0.05\,$MeV,
${\rm BR}(\phi\rightarrow\pi^{+}\pi^{-})=(7.3\pm1.3)\times10^{-5}$
from~\cite{R3} and 
${\rm BR}(\omega\rightarrow\pi^{+}\pi^{-})=(1.30\pm0.24\pm0.05)\%$
from~\cite{Akhmetshin:2003zn}. This gives for the on-shell mixing amplitude
\begin{eqnarray}
\Theta_{\rho\omega}=(-3.75\pm0.35\pm0.07)\times10^{-3}~{\rm GeV}^{2}~.
\end{eqnarray}
The uncertainty in the branching ratio of the process
$\omega\rightarrow\pi^{+}\pi^{-}$ causes the error in the value of
$\Theta_{\rho\omega}$ amplitude. Furthermore, we obtain  values of
$\Theta_{\phi\omega}$ that depend on $\Theta_{\rho\omega}$ (cf.
Eqs.~(\ref{ompipi},\ref{phipipi})):
\begin{eqnarray}
\Theta_{\phi\omega}= (25.34\pm2.39)\times10^{-3}~{\rm GeV}^{2}
\end{eqnarray}
Note that if we substitute
$M^{2}_{\omega}-M^{2}_{\rho}-i(M_{\omega}\Gamma_{\omega}
-M_{\rho}\Gamma_{\rho})$
with the dominant term $iM_{\rho}\Gamma_{\rho}$ 
in Eq.~(\ref{ompipi}), the $\rho-\omega$ mixing amplitude is
$\Theta_{\rho\omega}= (-3.96\pm0.37\pm0.08)\times10^{-3}~{\rm GeV}^{2}$
and the value affects the $\phi-\omega$ mixing amplitude as follows:
$\Theta_{\phi\omega}=(24.03\pm2.27)\times10^{-3}~{\rm GeV}^{2}\,$.
In Table~1 we have collected some values for the $\rho-\omega$ mixing 
amplitude \cite{Coon:1987kt,Bernicha:1994re,O'Connell:1995wf} (as extracted
from the pion vector form factor)
and the resulting $\omega-\phi$ mixing deduced from
Eqs.~(\ref{ompipi},\ref{phipipi}).
Our result of the $\rho-\omega$ mixing amplitude is good agreement
with these values. For further studies of $\rho$-$\omega$ and 
 $\rho$-$\omega$-$\phi$ mixing see \cite{Gardner:1997ie} and  
\cite{Benayoun:2001qz}, respectively.
Tab.~1 also includes the values of the $\phi-\omega$
amplitude which is calculated using the values of the
$\rho-\omega$ mixing amplitude. These values of the amplitude
$\Theta_{\phi\omega}$ (including ours) 
are consistent with the range  $20.00 \ldots 29.00\times10^{-3}\,
{\rm GeV}^{2}$ given in~\cite{Achasov:1989mh,Achasov:1989kz}. 
We can also give  the magnitude of the
$\phi-\omega$ mixing parameter defined in Eq.(\ref{defeps})
and the deviation from the ideal mixing angle,
\begin{equation}
\varepsilon~=~0.059\pm0.005~, \quad \varphi_{V}~=~3.4\pm0.3^{\circ}~,
\end{equation}
consistent with the findings e.g. in Refs.~\cite{Jain:1987sz,Bramon:1997va}.

\begin{table}[htb]
\caption{Values of the $\phi-\omega$ mixing amplitude that
depends on  the $\rho-\omega$ mixing amplitude
for various values from the literature in comparison to this work.}
\begin{center}
\begin{tabular}{|c|c|c|}
\hline Ref. & $\Theta_{\rho\omega}[\times10^{-3}~{\rm GeV}^{2}]$ &
$\Theta_{\phi\omega}[\times10^{-3}~{\rm GeV}^{2}]$ \\
\hline
\hline this work & $-3.75\pm0.36$ & $25.34\pm2.39$\\
\hline \cite{Urech:1995ry} & $-3.91 \pm 0.30$ & $24.31\pm1.88$ \\
\hline \cite{Coon:1987kt} & $-4.42\pm0.60$ & $21.03\pm2.84$\\
\hline \cite{Bernicha:1994re} & $-3.74\pm0.30$ & $25.41\pm2.05$\\
&$-4.23\pm0.68$ & $22.47\pm3.70$\\
&$-3.67\pm0.30$ & $25.90\pm2.13$\\
\hline \cite{O'Connell:1995wf} & $-3.97\pm0.20$ & $23.94\pm1.20$ \\
 \hline
\end{tabular}
\end{center}
\end{table}

\medskip
\noindent
From our analysis of $\rho$-$\omega$ mixing, we can update the value of the
quark mass ratio $R = (m_s - \hat m)/(m_d - m_u)$ using the formalism
developed in Ref.~\cite{Gasser:1982ap}. We find
\begin{equation}
R = 42 \pm 4~,
\end{equation}
where the uncertainty has been estimated in a similar fashion as 
in~\cite{Urech:1995ry}.


\medskip

\noindent {\bf 6.}
Finally, we consider the off-shell behavior of the two-point 
function, that allows to describe the momentum-dependence of the 
$\phi-\omega$ mixing.
Using the definition of the vector mesons in the tensor
representation
\begin{eqnarray}
V_{\mu}=\frac{1}{M_V} \,\partial^{\nu}V_{\mu\nu},
\end{eqnarray}
we obtain the following expression
\begin{eqnarray}
&i& \!\!\!\!  \int d^{4}x \, e^{ikx}\langle 0
|T\phi_{\mu}(x)\omega_{\nu}(0)e^{i\int d^{4}y\{{\cal
L}_{2}^{V}+{\cal L}_{2}^{\phi\omega}\}}|0\rangle \nonumber
\\
&=& \left(g_{\mu\nu}-\frac{k_{\mu}k_{\nu}}{k^{2}}\right)\frac{\Theta(k^{2})}
{(M_{\phi}^{2}-k^{2})(M_{\omega}^{2}-k^{2})}~,
\end{eqnarray}
with
\begin{equation}
\Theta(k^{2})=\left[2\sqrt{2}M_{\rho}(\widehat{m}-m_{s})+
\frac{\sqrt{2}}{9}e^{2}F_{V}^{2}\right]\frac{k^{2}}{M_{\phi}M_{\omega}}~.
\end{equation}
This result is similar to the one of Urech~\cite{Urech:1995ry} for the
momentum dependence of the $\rho-\omega$ mixing amplitude. Using similar
arguments than given in that paper, one concludes that the momentum dependence
of the amplitude for nucleon-nucleon scattering with resonance exchange
and $\omega$-$\phi$ mixing is more complicated than the one of $\Theta(k^{2})$
(we refrain from giving the pertinent formulae here).


\medskip

\noindent {\bf 7.}
In this note, we have considered $\omega$-$\phi$ mixing to leading order in
chiral perturbation theory with vector mesons and discussed some
implications. In the future, one should
consider loop corrections to these results utilizing either the heavy vector
meson chiral Lagrangian developed in~\cite{Jenkins:1995vb} or the infrared
regularization scheme for spin-1 fields presented in~\cite{Bruns:2004tj}.
Note that some of these issues were already addressed 
in~\cite{Bijnens:1996kg,Bijnens:1997ni}.

\acknowledgments{
We appreciate discussions  with Bugra Borasoy, Peter Bruns, J\"urg Gasser,
Hans-Werner Hammer, Bastian Kubis and Robin Ni{\ss}ler.
This work was supported in part  by Deutsche Forschungsgemeinschaft 
through funds provided to the SFB/TR 16 ``Subnuclear Structure of Matter''.
This research is part of the EU Integrated Infrastructure Initiative Hadron
Physics Project under contract number RII3-CT-2004-506078.
}

\end{document}